\documentclass[aps,pra,twocolumn,reprint,amssymb,floats,superscriptaddress]{revtex4}
\usepackage{array}
\usepackage{graphicx}   
\usepackage{color}      
\usepackage{amsmath}
\usepackage{listings}

\usepackage{color} 
\definecolor{mygreen}{RGB}{28,172,0} 
\definecolor{mylilas}{RGB}{170,55,241}

\usepackage{hyperref}   

\begin{document}

\title{From prethermalization to chaos in periodically driven coupled rotors}
\author{Yonathan Sadia}
\affiliation{Department of Physics, Bar-Ilan University, Ramat Gan 5290002, Israel}
\affiliation{Center for Quantum Entanglement Science and Technology, Bar-Ilan University, Ramat Gan 5290002, Israel}
\author{Emanuele G. Dalla Torre}
\affiliation{Department of Physics, Bar-Ilan University, Ramat Gan 5290002, Israel}
\affiliation{Center for Quantum Entanglement Science and Technology, Bar-Ilan University, Ramat Gan 5290002, Israel}
\author{Atanu Rajak}
\affiliation{Presidency University, Kolkata, West Bengal 700073, India}

\begin{abstract}
Periodically driven (Floquet) systems are said to prethermalize when their energy absorption is very slow for long time.
This effect was first discovered in quantum spin models, where the heating rate is exponentially small in the ratio between the driving frequency and the spin bandwidth. Recently, it was shown that prethermalization occurs also in classical  systems with an infinite bandwidth. Here, we address the open question of which small parameter controls the lifetime of the prethermal state in these systems. We, first, numerically study the dependence of the lifetime on the initial conditions and on the connectivity in a system of periodically driven coupled rotors. We find that the lifetime is controlled by the temperature of the prethermal state, which is quasi-conserved when the heating is slow.  This finding allows us to develop a simple analytical model that describes the crossover from prethermalization to chaos in many-body classical systems.
\end{abstract}

\maketitle


\section{Introduction}
Periodic driving of isolated many-body systems can generate novel dynamical phases that do not have any static analogues. This approach, known as Floquet engineering \cite{oka2019floquet,rudner2020band,weitenberg2021tailoring}, led to a plethora of new phases, such as (anomalous) Floquet topological insulator~\cite{oka09photovoltaic,lindner11floquet,rechtsman2013photonic,katan2013modulated,jotzu14experimental,fleury2016floquet,peng2016experimental,roy2017periodic,maczewsky2017observation,wintersperger2020realization,mukherjee2020observation}
and time crystals~\cite{khemani16phase,else16floquet,zhang17observation,yao17discrete,russomanno2017floquet,moessner2017equilibration,rovny2018observation}.
However, periodic driving also introduces heating, which hinders such novel applications. 
The heating can be suppressed by introducing strong disorder in interacting many-body systems,
thus creating many-body localized phases \cite{ponte15many,lazarides15fate}. The drawback of this method
is that the disorder can destroy the generic characteristics of the non-equilibrium phases. An alternative method that can be used to reduce heating is
driving the systems at high frequencies, leading to a long-lived prethermal state, where the heating rate is exponentially 
slowed down. This phenomenon was first discovered in quantum many-body systems 
~\cite{choudhury14stability,bukov15prethermal,goldman15periodically,chandran16interaction,
	lellouch17parametric,lellouch2018parametric,abanin15exponentially,kuwahara16floquet,mori2016rigorous,abanin17rigorous,gulden2020exponentially}. 
Recently, using quantum simulator~\cite{rubio2020floquet,kyprianidis2021observation} and NMR techniques~\cite{peng2021floquet}, the existence of long-lived prethermal states 
at high frequency driving has been observed in experiments. 
The existence of prethermal plateau has also been observed in interacting quantum kicked rotor, realized experimentally in optical lattices~\cite{cao21prethermal}.

A fundamental question is whether the phenomenon of Floquet prethermalization can be found in classical systems as well. The answer to this question has been given affirmatively in the recent 
literature~\cite{citro2015dynamical,rajak2018stability,howell2019asymptotic,mori2018floquet,rajak2019characterizations,torre2021statistical}.
Using canonical models of classical chaos theory~\cite{rajak2018stability,rajak2019characterizations} and classical 
driven spin chains~\cite{howell2019asymptotic,mori2018floquet}, a quasistationary prethermal regime 
has been found, before heating begins. In contrast to quantum systems, where Floquet prethermalization rigorously applies to models with  bounded operators, classical prethermalization occurs in systems with unbounded spectra and has a statistical nature~\cite{rajak2018stability,rajak2019characterizations}.

The prethermal state of periodically driven classical chaotic systems can be characterized by a generalized Gibbs ensemble (GGE)~\cite{rajak2019characterizations,torre2021statistical,das2020nonlinear}. For example, in the case of coupled rotors, the total angular momentum is a true conserved quantity, whereas the energy is quasi-conserved inside the prethermal regime. 
The temperature of the prethermal state can be calculated by equating the energy of the initial ensemble, with the average energy of the GGE. 
When the ratio between the driving frequency and the temperature is large, the heating rate is suppressed by the low probability for the GGE to satisfy the conditions of a many-body resonance \cite{rajak2019characterizations}, leading to a statistical Floquet prethermalization. In a recent work, the different dynamical regimes of the system were further characterized by considering spatio-temporal correlations~\cite{kundu2021dynamics}. In analogy to the static case~\cite{das2020nonlinear}, these 
correlations show a diffusive behavior inside the prethermal regime, thus supporting the quasi-static nature of 
the prethermal state.

In this article, we consider the effect of initial conditions 
and connectivity of the rotors on the lifetime of the prethermal states. Our main result is that the two effects act in a similar way, namely by affecting the initial energy and, hence, changing the temperature of the prethermal state. First, the effect of initial conditions is investigated by tuning the standard deviation of the angles of the rotors in the initial state.
By studying the dependence of the lifetime of the prethermal state on the standard deviation, we 
establish that the lifetime depends exponentially on the inverse temperature of the prethermal state. 
Next, we investigate the connectivity of the rotors by considering a many-body kicked rotor model where all the rotors interact with each other. Unlike the nearest-neighbor case, we find that the kinetic energy 
per rotor, for fixed initial conditions, depends on the number of rotors $N$, and the prethermal temperature 
increases linearly with $\sqrt{N}$. Also in this case, the lifetime varies exponentially as the inverse temperature. 
Starting from these results, we propose an analytical {\it ansatz} that describes the universal properties of the crossover from the prethermal regime to the chaotic one.


\section{The model: coupled kicked rotors in one and higher dimensions}
\label{model}
In this work we consider a canonical example of chaotic, classical, many-body systems, namely the coupled kicked rotors Hamiltonian
~\cite{chirikov79universal,kaneko89diffusion,konishi90diffusion,chirikov1993theory,antoni95clustering,mulansky11strong,rajak2020stability}  
\begin{align}
H(t)=\sum_{i=1}^{N}\frac{p_i^2}{2} - \Delta(t)\sum_{i,j=1,i<j}^N\kappa_{i,j}\cos(\phi_{i}-\phi_{j}).
\label{ham_system}
\end{align}
Here, $p_i$ and $\phi_i$ are, respectively, the angular momentum and the angle of $i$-th rotor and $N$ is 
the total number of rotors.
The system is periodically driven with delta-function kicks, 
$\Delta(t)=\sum_n\delta(t-n\tau)$, where $\tau$ is the time period. The couplings $\kappa_{i,j}$ correspond to
the interactions between the rotors and define the kick strength. We consider two types 
of interactions between the rotors: (i) a one-dimensional model with nearest-neighbor coupling, where $\kappa_{i,j}=\kappa(\delta_{i,j-1}+\delta_{i,j+1})$, and (ii) 
a mean-field model with
all-to-all coupling, where $\kappa_{i,j}=\kappa/\sqrt{N}$ \footnote{The normalization of the latter model is discussed below.}.

Using Hamilton's equations of motion, one obtains a discrete map of $p_i$ and $\phi_i$ between consecutive kicks,
\begin{align}
\begin{split}
p_i(t+\tau)&=p_i(t)-\sum_{j\neq i}\kappa_{i,j}\sin(\phi_i-\phi_j),\\
\phi_i(t+\tau)&=\phi_i(t)+p_i(t+\tau)\tau.
\end{split}
\label{eq_motion}
\end{align}
These equations are a many-body generalization of Chirikov standard map \cite{chirikov2008chirikov}, a system of paramount importance for the study of the transition between regular and chaotic dynamics. By rescaling $p_i\to p_i\tau$, one finds that the equations of motion (\ref{eq_motion}) are characterized by a single dimensionless parameter, $K=\kappa\tau$. Hence, in what follows, we will set $\tau=1$ without loss of generality. As we will see, the initial conditions and the connectivity introduce additional unitless parameters that can be used to tune the prethermal regime. 






Because the model (\ref{ham_system}) is non-integrable, at long times it shows a chaotic behavior, where the kinetic energy $E_{\rm kin,i}(t)=\langle p_i^2\rangle /2$ grows linearly with time, $E_{\rm kin,i}(t) \approx \Gamma t$ \footnote{The heating rate $\Gamma$ is related to the diffusion coefficient $D$, defined as $\langle p_i^2\rangle=Dt$, by $\Gamma=D/2$}. The heating rate $\Gamma$ can be estimated using the following approach: According to Eqs.~(\ref{eq_motion}), the momentum at time $t$ equals to
\begin{align}
p_{i}(t)=p_i(0)-\sum_{n=0}^{t-1}\sum_{j\neq i} \kappa_{i,j} \sin(\phi_i(n)-\phi_j(n))
\label{momentum_time}
\end{align}
Now, squaring Eq.~(\ref{momentum_time}) and averaging over symmetric initial conditions with $\langle p_i(0)\rangle=0$, we obtain 
\begin{align}
&\langle p^2_{i}(t)\rangle=\langle p_i^2(0)\rangle+\label{eq:p2_mean_field}\\
&\sum_{m,n=0}^{t-1}\sum_{j,k\neq i}\kappa_{i,j}\kappa_{i,k} \Big\langle \sin(\phi_i(m)-\phi_j(m))\sin(\phi_i(n)-\phi_k(n))\Big\rangle.
\nonumber\end{align}
In the chaotic regime, the angles of the rotors become statistically uncorrelated both in space and time, and we can approximate
$\left\langle\sin(\phi_i(m)-\phi_j(m))\sin(\phi_i(n)-\phi_k(n))\right\rangle=\delta_{j,k}\delta_{m,n}/2$ 

\begin{figure}[t]
\includegraphics[width=\linewidth]{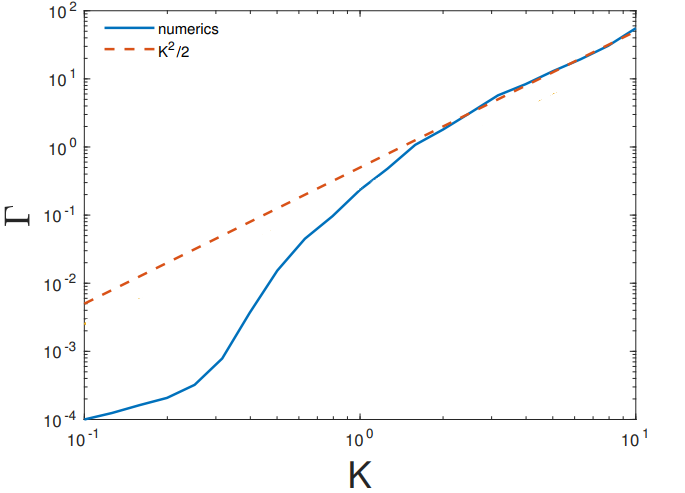}
\caption{Heating rate for one-dimensional model, as a function of the coupling parameter $K$. For large $K$, it follows the relation $K^2/2$.}
\label{fig:1d_diffusion}
\end{figure}

For the one-dimensional model, only the terms $j=k=i\pm 1$ contribute to the sum of Eq.~(\ref{eq:p2_mean_field}), 
leading to $\langle p^2_i\rangle = K^2 t$, or equivalently  $\Gamma=K^2/2$. From the comparison with numerical simulations of the model, 
it was found that our analytical result for the heating rate is valid for large $K$ (see Fig.~\ref{fig:1d_diffusion}). For small $K$, the rotors move in a correlated way and, for 
$K\lesssim 0.1$ the heating is approximately given by $\Gamma\approx 10 K^{6.5}$ \cite{konishi90diffusion,chirikov1993theory,mulansky11strong}. 
This effect, known as ``fast Arnold diffusion'' was explained in Refs.  \cite{chirikov1993theory,chirikov97arnold} using the concept of 
many-body resonance \cite{chirikov79universal}. Importantly, for all values of $K$, the heating rate does not 
depend on $N$ \footnote{see Appendix B of Ref.~\cite{rajak2018stability}}. 

\begin{figure}[t]
	\centering
	(a) all-to-all coupling: small $K$\\
	\includegraphics[width=\linewidth]{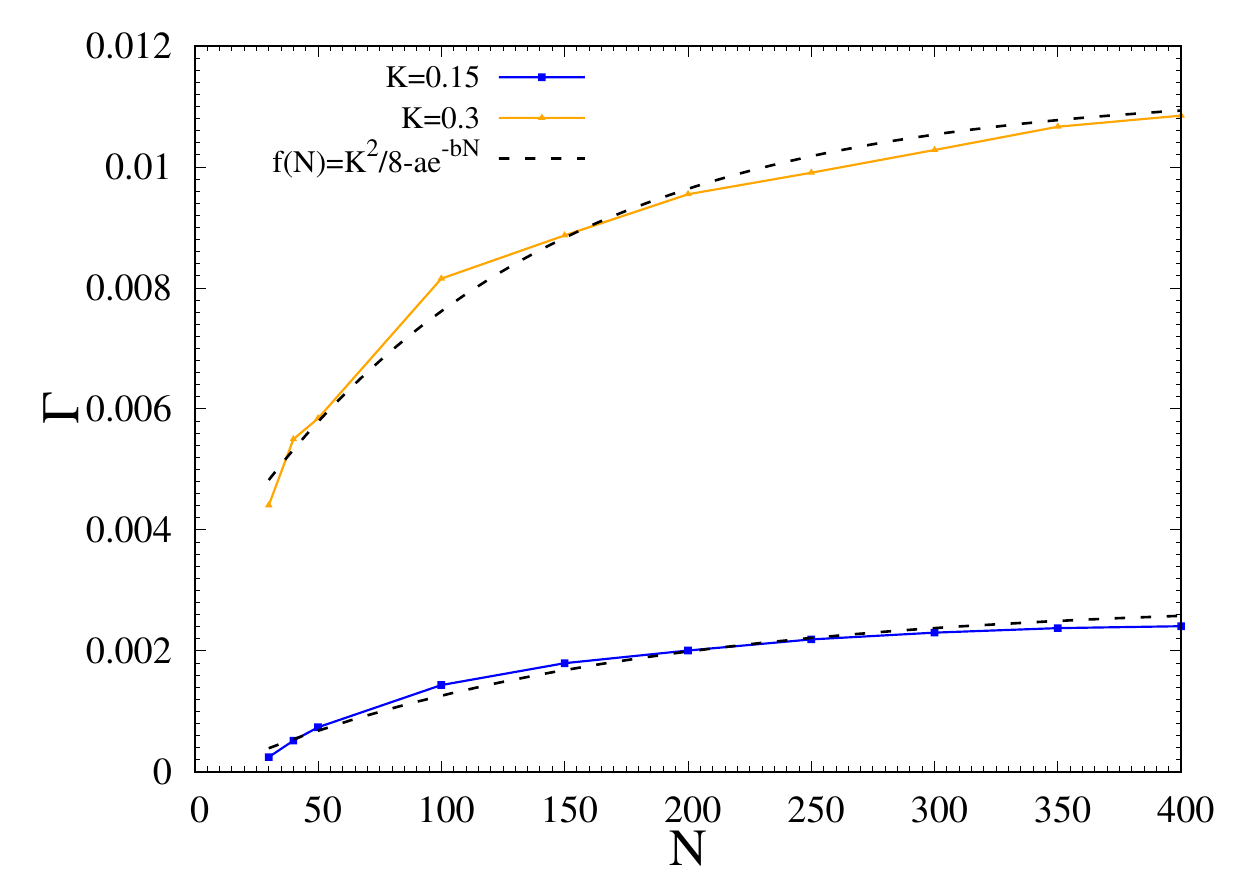}\\
	(b) all-to-all coupling: large $K$\\
	\includegraphics[width=\linewidth]{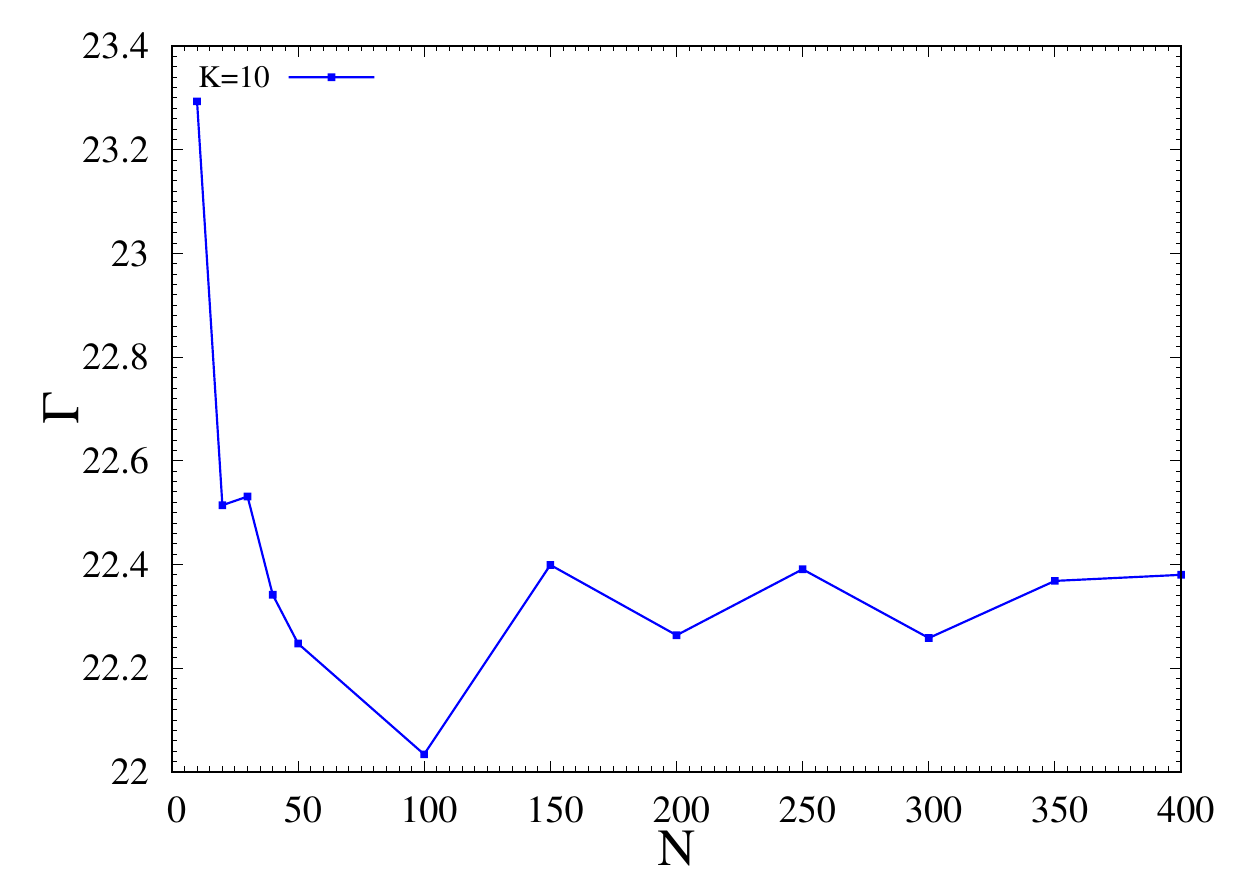}
	\caption{Heating rate as a function of $N$ of all-to-all coupling model for (a) small values of $K$ and (b) a large value of $K$.}
	\label{fig:atanu_diffusion}
\end{figure}

In the case of all-to-all coupling, all terms with $j=k\neq i$ equally contribute to Eq.~(\ref{eq:p2_mean_field}) and one obtains 
$\langle p^2_i(t)\rangle =  (N-1)\kappa^2/(2N)$. For large $N$, the heating rate, $\Gamma \approx K^2/4$, becomes independent on $N$. 
To verify the validity of the uncorrelated behavior of the rotors in the heating regime, we numerically compute the dynamics of the system for long times and compute 
the heating rate in the chaotic regime, see Fig.~\ref{fig:atanu_diffusion}. For small values of the coupling parameter, say $K=0.15, 0.3$,
we find that $\Gamma$ increases with $N$ and saturates to a value $K^2/8$ for large $N$ (see Fig.~\ref{fig:atanu_diffusion}(a)), whereas the theoretically 
expected value is $K^2/4$. This anomaly is resolved by plotting $\Gamma$ as a function of $N$ for large $K$. 
From Fig.~\ref{fig:atanu_diffusion}(b), we see that the heating rate saturates to a value closer to $K^2/4$ for large $N$ 
with a large $K$, i.e., when the system is well inside the chaotic regime. 

These results indicate that in both models, our assumption of an uncorrelated behavior of the angles is valid for large values of $K$ only. 
For small values of $K$ the rotors undergo a correlated dynamics, even in the diffusive long-time limit. Importantly, our simulations demonstrate 
that for all values of $K$, in both models the heating rate does not depend on $N$ for large enough $N$.


\section{Energy and temperature in the prethermal state}
\label{prethermal_state}
We open our discussion by considering the prethermal regime, shortly after the initial conditions. In systems displaying statistical Floquet prethermalization, the time-averaged Hamiltonian $H_{\rm av}$ is quasi-conserved. Here, 
	\begin{align}
	H_{\rm av} = \frac1{\tau}\int_0^\tau dt~H(t) = \sum_{i=1}^{N}\frac{p_i^2}{2} - \sum_{i<j}^N\frac{\kappa_{ij}}{\tau}\cos(\phi_{i}-\phi_{j}).
	\label{ham_avrg}
	\end{align} 
At short times, we can assume that the average energy in the prethermal state equals to the average energy in the initial state
\begin{align}
\langle H_{\rm av} \rangle_T = \langle H_{\rm av} \rangle_0, \label{eq:HTH0}
\end{align}
where $\langle ...\rangle_T$ is the Boltzmann distribution with Hamiltonian $H_{\rm av}$ and temperature $T$, and $\langle ...\rangle_0$ is the average over the initial conditions. 
This equation can be used to derive the temperature of the prethermal state from a given set of initial conditions.

\begin{figure}[t]
	\centering
	(a) one dimension\\
	\includegraphics[width=\linewidth]{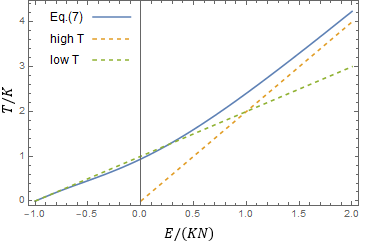}\\
	(b) all-to-all coupling\\
	\includegraphics[width=\linewidth]{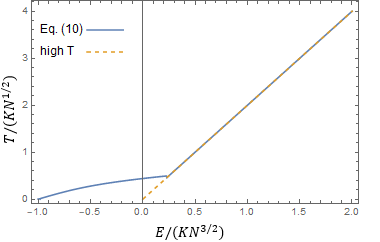}
	\caption{Temperature as a function of the energy $E=\langle H_{\rm av}\rangle$ for (a) the one dimensional model and (b) the all-to-all coupling model, obtained by the numerical solution of Eq.~(\ref{eq:1d}) and (\ref{eq:mf}), respectively. The dashed lines are asymptotic results at small and large energies (see text).}
	\label{fig:temperature}
\end{figure}

\begin{figure}[t]
	\centering
	(a) one dimension\\
	\includegraphics[width=\linewidth]{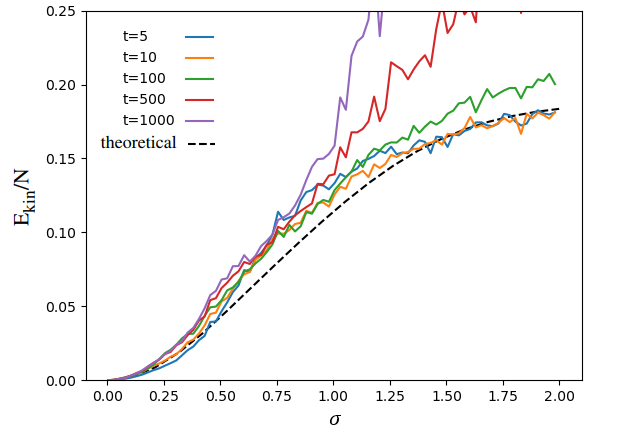}
	(b) all-to-all coupling\\
	\includegraphics[width=\linewidth]{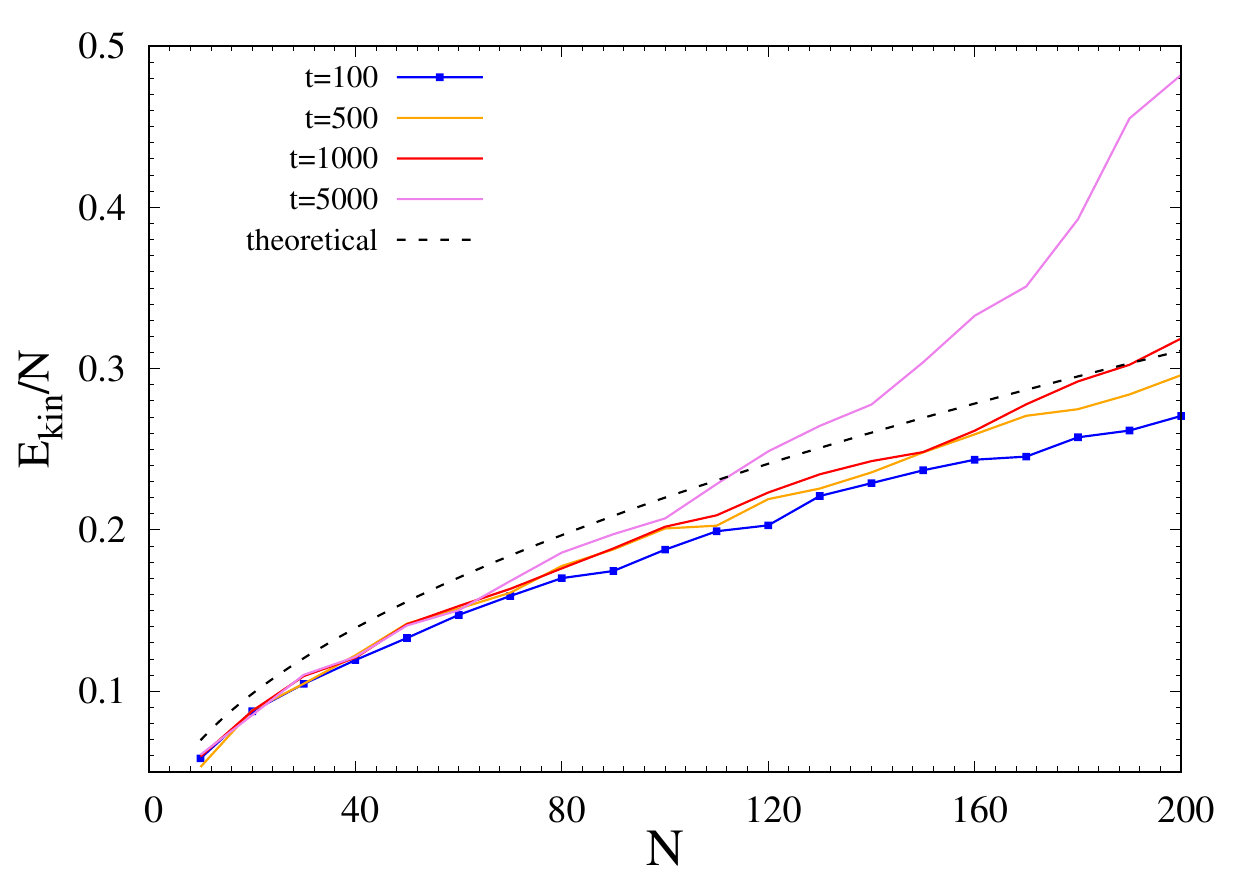}
	\caption{Kinetic energy per rotor in the prethermal state: (a) in the one-dimensional model for $K=0.4$, as a function of the initial fluctuations $\sigma$; (b) in the all-to-all coupling model for $K=0.1$, as a function of the number of rotors $N$. The dashed lines are our analytical predictions (see text for details).}
	
	\label{fig:prethermal_energy}
\end{figure}

Let us, first, consider the one dimensional model, where the thermal average can be performed exactly \cite{rajak2019characterizations} and leads to
\begin{align}
\frac{\langle H_{\rm av}\rangle_{T}}N = \frac{T}2 - K \frac{I_1(K/T)}{I_0(K/T)}\label{eq:1d}
\end{align} 
Here, the right-hand side corresponds to the sum of the kinetic energy per rotor, $T/2$, and the potential energy per rotor, expressed in terms of the modified Bessel functions $I_0$ and $I_1$. The numerical solution of this equation is shown in Fig.~\ref{fig:temperature}(a). At large temperatures, $T\gg K$, we can neglect the potential energy and obtain $\langle H_{\rm av}\rangle_T = NT/2$. In the opposite limit, $T\ll K$, we find $\langle H_{\rm av}\rangle = N(T-K)$. 
This result can be understood by observing that at small temperatures we can approximate $\cos(\phi)\approx 1-\phi^2/2$, leading to a set of harmonic oscillators, with average potential energy $-K+T/2$. The situation studied in Ref.~\cite{rajak2019characterizations} corresponds to the case $E=0$, where $T=0.9384K$. 

In our numerical simulations we consider $p_i(t=0)=0$ and extract $\phi_i(t=0)$ from a Gaussian distribution with standard deviation $\sigma$. In this case, the initial energy is
\begin{align}\frac{\langle H_{\rm av}\rangle_0}N = - K e^{-\sigma^2},\label{eq:E01d}\end{align} 
The free parameter $\sigma$ controls the temperature of the initial state and, consequently, the energy of the prethermal state. By equating Eq.~(\ref{eq:E01d}) and Eq.~(\ref{eq:1d}), we find an implicit relation between the parameter $\sigma$ and the temperature of the prethermal state. In Fig.~\ref{fig:prethermal_energy}(a) we show that the predicted kinetic energy $E_{\rm kin}/N=T/2$ exactly matches the numerical solution of the model.
For $\sigma\to\infty$, we recover the result of Ref.~\cite{rajak2019characterizations} $T=0.9384K$.

In the case of all-to-all coupling, the dependence between the energy and the temperature can be computed within a mean-field approximation \cite{antoni95clustering}
\begin{align}
\sum_j\cos(\phi_i-\phi_{j}) \approx N(\langle\cos\phi\rangle \cos\phi_i+\langle\sin\phi\rangle \sin\phi_i),\label{eq:meanfield}
\end{align}
such that
\begin{align}
\frac{\langle H_{\rm av}\rangle _T}N = \frac{T}2 - \frac{\kappa}{2\tau} N^{1/2} \left(\langle \cos\phi\rangle^2 + \langle \sin\phi\rangle^2\right),\label{eq:mf}
\end{align}
with
\begin{align}
&\langle\cos\phi\rangle = \langle\sin\phi\rangle = \frac{I_1(c)}{I_0(c)}~~~{\rm and}~~~ c = \frac{K\sqrt{N}}{2 T}\langle \cos\phi\rangle.\nonumber
\end{align} 
The last line corresponds to a Boltzmann average over the Hamiltonian $H_{\rm av}$, with the approximation (\ref{eq:meanfield}). Note that Eq.~(\ref{eq:mf}) is a function of the rescaled parameters $E/(KN^{3/2})$ and $T/(K\sqrt{N})$ only. The numerical solution of this equation in shown in Fig.~\ref{fig:temperature}(b). At high temperatures $\langle \cos(\phi)\rangle=0$ and one simply has $E=NT/2$ (the potential energy becomes negligible). 
%
%
In our numerical simulations the rotors are initialized at random angle between $0$ and $2\pi$, with $p_i(t=0)=0$, such that the initial energy is $\langle H_{\rm av}\rangle _0=0$ and the temperature of the prethermal state is $T= 0.44 K\sqrt{N}$, see Fig.~\ref{fig:prethermal_energy}(b). Hence, in the all-to-all coupling model the initial energy and the temperature of the prethermal state are controlled by $N$.


\begin{figure}[ht!]
	\includegraphics[width=\linewidth]{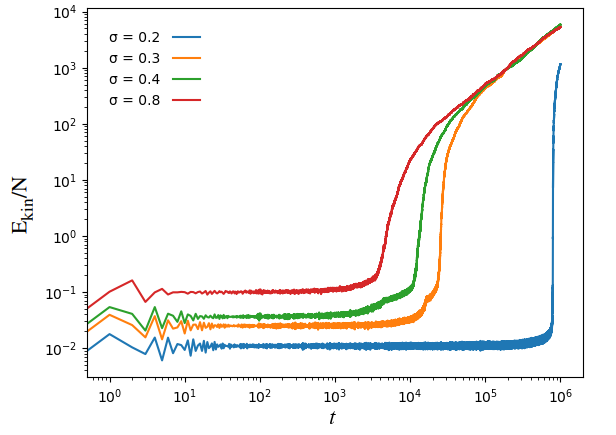}
	\caption{Time evolution of the kinetic energy for the one-dimensional model ($K=0.4$), for different values of the standard deviation of initial angles, $\sigma$. As shown in Fig.~\ref{fig:yonathan2}, all the curves correspond to the same function, shifted in the time axis. The apparent change of the slope is an artifact of the logarithmic scale.}
	\label{fig:yonathan1}
	\includegraphics[width=0.9\linewidth]{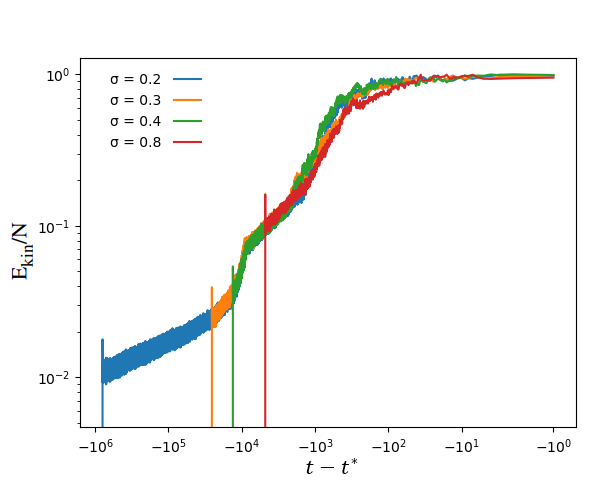} \\
	\includegraphics[width=0.9\linewidth]{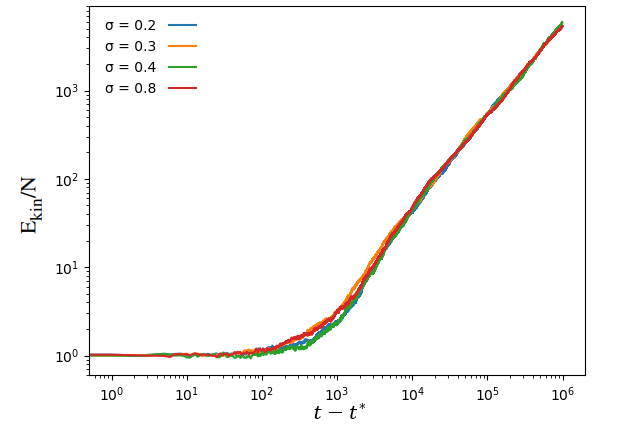}
	\caption{Same curves as Fig.~\ref{fig:yonathan1}, plotted as a function of  $t^*-t$, for $t<t^*$ (upper panel), and $t-t^*$ for $t>t^*$ (lower panel). The excellent data collapse demonstrates that the transition from the prethermal regime to the chaotic regime is universal and does not depend on the initial conditions.}
	\label{fig:yonathan2}
\end{figure}

\section{From prethermalization to chaos}
\label{pre_diff}
 \subsection{Numerics in one dimension}
 \label{sec:1d}
\begin{figure}[t]
	\includegraphics[width=\linewidth]{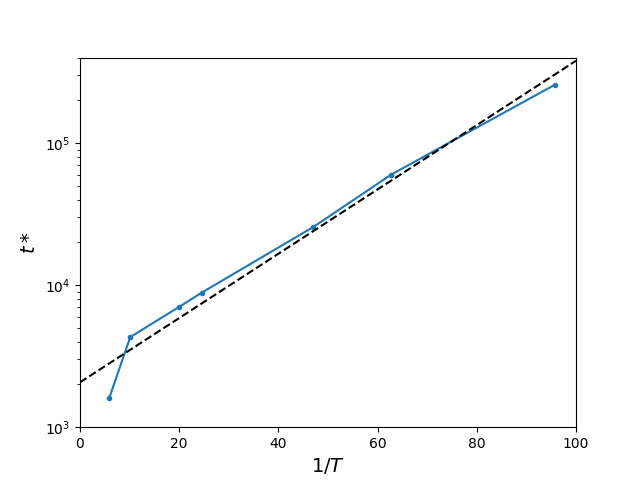}
	\caption{Lifetime of the prethermal state of the one-dimensional model with $K=0.4$, as a function of the inverse temperature of the prethermal state $1/T$, obtained by varying $\sigma$. The lifetime is exponentially large in the inverse temperature.}
	\label{fig:yonathan3}
\end{figure}

 We now focus on the transition between the prethermal state and the chaotic regime, starting from the one-dimensional model. 
 Figure~\ref{fig:yonathan1} shows the time evolution of the kinetic energy per rotor for different values of the initial fluctuations' parameter $\sigma$.  Note that, in this plot, the transition between the prethermal and chaotic regimes appear to become sharper with decreasing $\sigma$. This is inconsistent with the normalization procedure proposed by Ref.~\cite{rajak2018stability},
 $t\to t/t^*$, which corresponds to a rigid shift in the logarithmic scale. Interestingly, we observed that the curves collapse over many orders of magnitude, when a rigid shifted is applied on a linear scale, $t\to t-t^*$. To demonstrate this effect, in Fig.~\ref{fig:yonathan2} we define $t^*$ by $E_{\rm kin}(t^*)/N=1$ and plot $E_{
 \rm kin}(t-t^*)$, obtaining a perfect data-collapse for both $t<t^*$ and $t>t^*$. In Fig.~\ref{fig:yonathan3} we show the dependence of $t^*$ on the inverse temperature of the prethermal state and find an exponential behavior. These numerical findings will be explained by the analytical model developed in Sec.~\ref{sec:ansatz}.

\subsection{Numerics of all-to-all coupling}
\label{alltoall}
\begin{figure}[t]
	\includegraphics[scale=0.6]{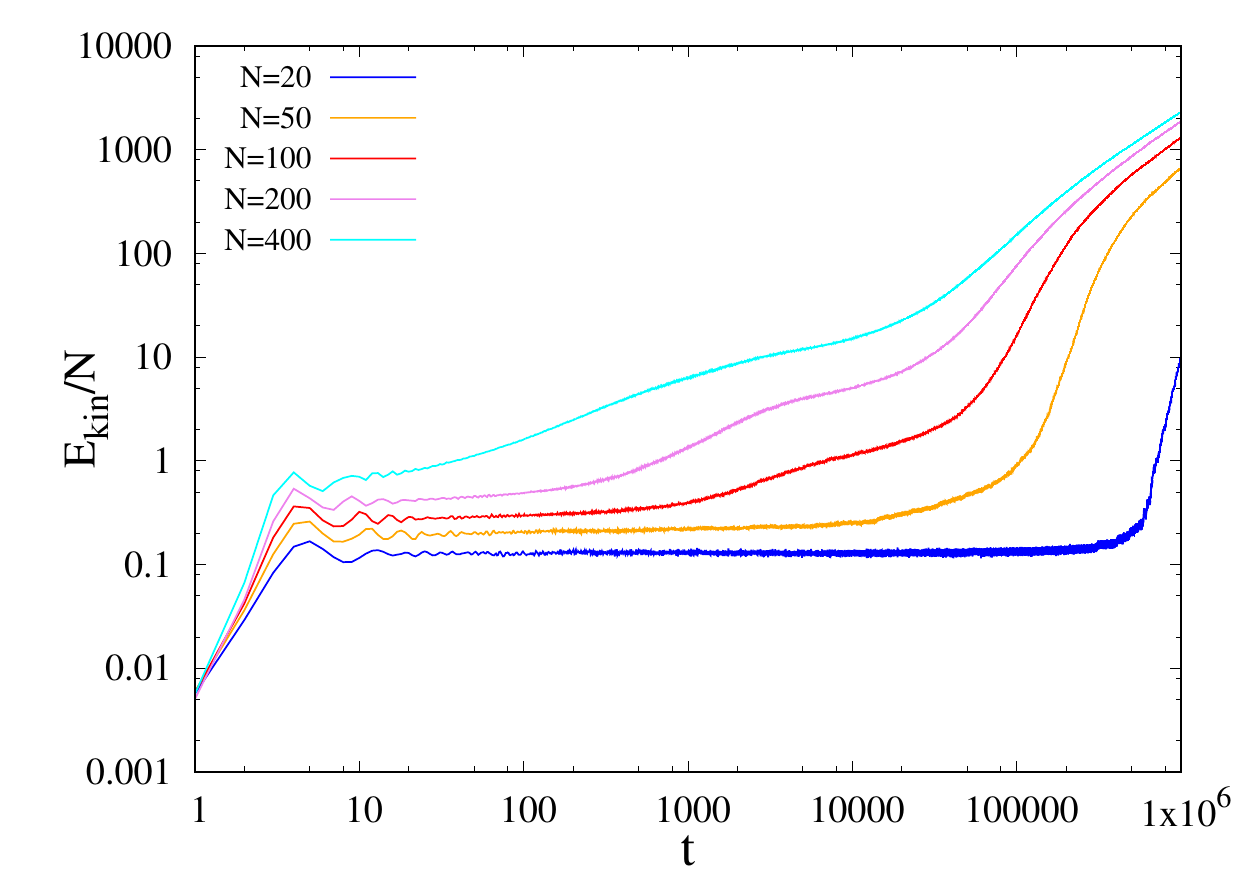}
	\caption{Time-evolution of the kinetic energy per rotor for the all-to-all coupling with $K=0.15$, for different values of the number of rotors $N$. The number of rotors affects the temperature of the prethermal state and its lifetime, while the heating rate in the chaotic regime remains constant.}
 	\label{fig:atanu_time}
 \end{figure}

We now move to the case of all-to-all coupling. In Sec.~\ref{model}, we used a mean-field theory to compute the heating rate in the chaotic regime and demonstrated that it does not depend on number of rotors $N$. In Sec.~\ref{prethermal_state} we showed that the temperature of the prethermal state increases as $T=0.44K\sqrt{N}$. Hence, we expect that as we increase $N$, the lifetime of the prethermal state should decrease. This behavior is indeed observed in the numerical solution of the model, see Fig.~\ref{fig:atanu_time}, 
To explore this effect in a quantitative manner, we plot the  lifetime of the prethermal states (defined by $E_{\rm kin}(t^*)/N=1$), as a function of the inverse temperature, and observe two distinct regimes: (i) small inverse temperatures $1/T<2.5$, corresponding to large number of rotors $N>(0.4/0.44K)^2\approx 36$; (ii) large inverse temperatures $1/T>2.5$, corresponding to small number of rotors ($N<36$). In the former regime, the heating rate is approximately constant and we observe an exponential suppression of heating. In the latter, finite-size effects are significant (see Fig.~\ref{fig:atanu_diffusion}) and cause a further suppression of heating.

 \begin{figure}[t]
	\includegraphics[width=\linewidth]{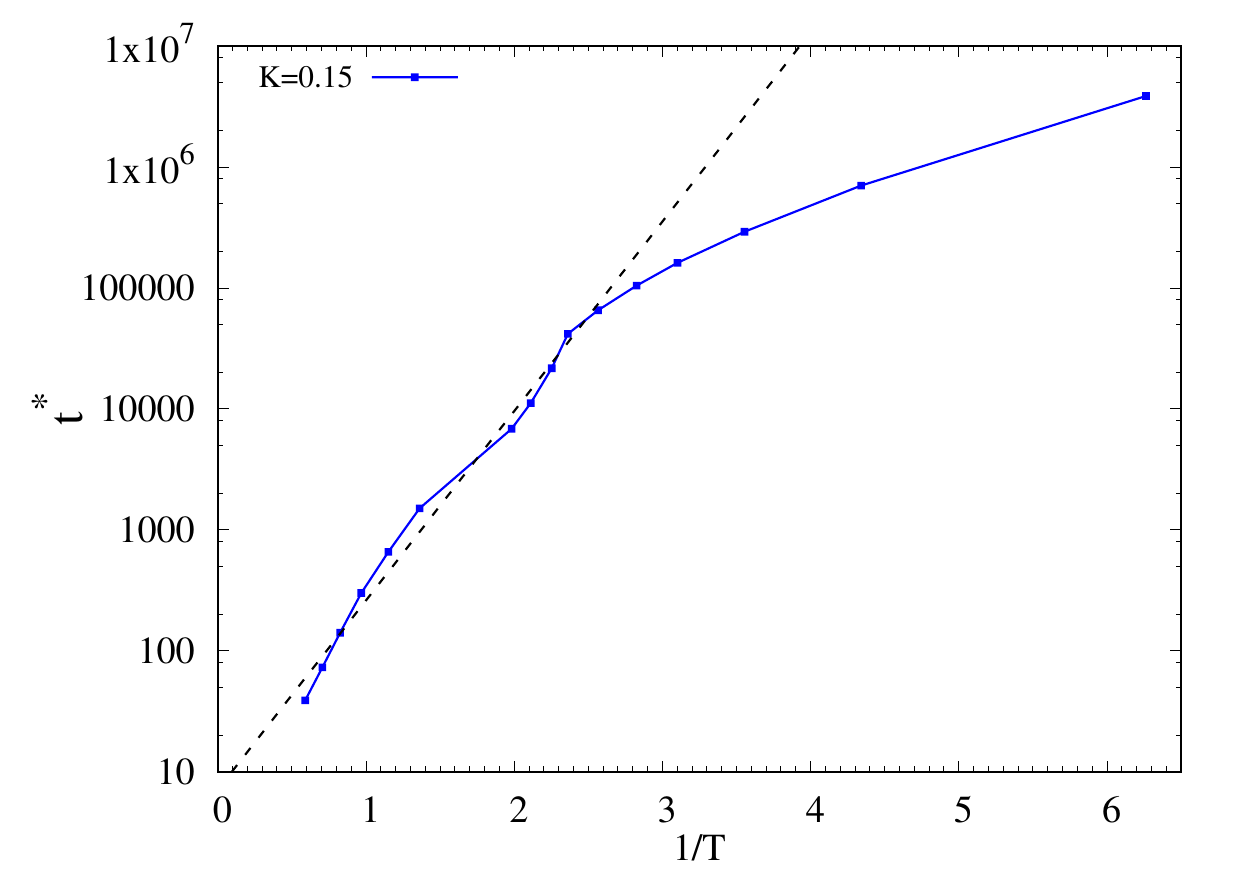}
	\caption{Lifetime of the prethermal state for the all-to-all coupling with $K=0.15$, as a function of the inverse temperature of the prethermal state. 
	The temperature $T$ is calculated numerically by $T=2E_{\rm kin}/N$, where $E_{\rm kin}$ is the kinetic energy in the prethermal state and $T$ is varied 
	by changing $N$ between $10$ and $400$. The dashed line is an exponential fit of the large temperature regime (see text).}
 	\label{fig:atanu_lifetime}
 \end{figure}

\subsection{Effective analytic description}
\label{sec:ansatz}

We now present a simple model that describes the escape from the prethermal regime to the chaotic one. The key assumption of the model is that, due to the exponentially slow absorption of energy, the prethermal state is a quasi-equilibrium state, characterized by an instantanous temperature $T(t)$. We consider the generic situation where the energy absorption depends exponentially on the temperature as
\begin{align}
\frac{dE(t)}{dt} = N \Gamma e^{-A/T(t)}\label{eq:dEdt}
\end{align}
Here $\Gamma=\Gamma(K)$ is the heating rate in the high-temperature, chaotic regime, where $dE/dt=N\Gamma$. In models of kicked rotors, this exponential suppression of heating is due to the low probability to find a rotor with angular momentum $p_i\sim\Omega=2\pi/\tau$ in a Boltzmann-Gibbs distribution with temperature $T(t)\ll K$ \cite{rajak2019characterizations}. See also Ref.~\cite{torre2021statistical} for the case of the Bose-Hubbard model, where the exponential suppression is associated to the low probability to find sites with large occupation numbers.

\begin{figure}[t!]
	\includegraphics[width=\linewidth]{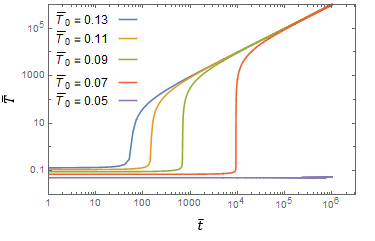}
	\caption{Time evolution of the temperature, obtained by the numerical solution of the rescaled analytical model, Eq.~(\ref{eq:rescaled}), for different initial conditions. The dashed lines are the lifetimes of the prethermal state predicted by Eq.~(\ref{eq:tstar}).}
	\label{fig:rescaled}
\end{figure}

In order to solve Eq.~(\ref{eq:dEdt}), we need to combine the energy-temperature relation $T=T(E)$. As shown in Sec.~\ref{prethermal_state}, this relation is often linear. For simplicity, we use here the high-temperature result $T=2E/N$, such that
\begin{align}
\frac{dT}{dt} = 2\Gamma e^{-A/T(t)}\label{eq:dTdt}
\end{align}
At short times, $T(t)\approx (T_0 + 2\Gamma e^{-A/T_0} dt)$ and the energy absorption is exponentially suppressed, leading to a long-lived prethermal regime. In contrast, at long times $T(t)\approx 2\Gamma t$, corresponding to the chaotic regime. Equation (\ref{eq:dTdt}) offers the minimal model of statistical prethermalization, where an exponentially long prethermal regime, is followed by a chaotic regime with linearly increasing temperature.

To compute the solution at intermediate times, it is useful to introduce the rescaled temperatures $\bar{T}=T/A$ and time $\bar{t}=2\Gamma t/A$, satisfying
\begin{align}
\frac{d\bar{T}}{d\bar{t}} = e^{-1/\bar{T}}\label{eq:rescaled}.
\end{align}
This equation has an implicit solution in terms of incomplete Gamma functions, shown in Fig.~(\ref{fig:rescaled}) for different values of $\bar{T}(0)$. Note that all the curves are identical, up to a shift in the $\bar{t}$ axis and the increasing sharpness for decreasing $\bar{T_0}$ is an artifact of the logarithmic scale (see Sect.~\ref{sec:1d}).


We can evaluate the lifetime of the prethermal, by computing the time at which the curve reaches some target value. This is equivalent to solving the inverse of Eq.~(\ref{eq:rescaled}), namely
\begin{align}
t^* = \int_{\bar{T}_0}^1 d{\bar T} e^{1/\bar{T}} \label{eq:deltat}
\end{align}
Here we have set the upper limit to $\bar{T}=1$, such that $t^*$ is defined as the time required to reach $\bar{T}(t^*)=1$. The integral in Eq.~(\ref{eq:deltat}) is readily solved to deliver
\begin{align}
t^* ={\rm Ei}\left(\frac{1}{\bar{T_0}}\right)-\bar{T}_0 \exp\left(\frac1{\bar{T}_0}\right) - {\rm Ei}(1) + e,
\label{eq:tstar}
\end{align}
where ${\rm Ei}$ is an exponential integral function. For large $x$, one has ${\rm Ei}(x)\approx (1/x+1/x^2)e^{x}$, leading to
\begin{align}
t^* = \bar{T}_0^2 \exp\left(\frac1{\bar{T}_0}\right).
\label{eq:tstar2}
\end{align}
This result shows that the lifetime of the prethermal time depends exponentially upon the inverse temperature of the state itself, as seen in our numerical solution of the one dimensional and all-to-all coupling models.

\section{Conclusion}
\label{conclu}
In conclusion, we performed a detailed study of the transition from prethermalization to chaos in classical periodically driven systems. 
We considered two tuning parameters that affect the temperature of the prethermal state, without changing the heating rate. The role of initial conditions is studied by a one-dimensional model where the temperature is set by the standard deviation of the initial Gaussian distribution of the angles. The effect of connectivity is studied by a system where all the rotors interact with each other and the temperature is a function of the number of rotors. In both cases, we computed the lifetime of the prethermal state and found that it depends exponentially on the inverse of the prethermal temperature. We repeated the same calculations in two and three dimensional lattice (not reported here), delivering similar results.

Starting from these numerical results, we proposed a simple model that describes the transition between prethermalization and chaos. Our model relies on two intertwined assumptions, namely that the prethermal state is fully described by its instantaneous temperature and that the heating rate is exponentially suppressed at low temperatures. The analytical solution of the resulting differential equation shows the same qualitative behavior as the numerical calculations. In particular, the lifetime of the prethermal state depends exponentially on the initial temperature of the prethermal state, see Eq.~(\ref{eq:tstar2}). In addition, our analytical model predicts that the curves for different initial conditions can be collapsed by a rigid shift in the time domain, as indeed observed numerically.

Our work rises several questions related to the relation between statistical Floquet prethermalization and other fundamental properties of chaotic systems.
The exponential scaling of the heating rate has some similarity with the phenomenon of Arnol'd diffusion, characteristic of systems close to integrability. In these systems, according to the Nekhoroshev theorem, the diffusion rate is bounded by an exponential function of $1/\epsilon^b$, where  $\epsilon$ is the perturbation from integrability and $b$ depends on the system size~\cite{nnn}. In spite of the similarity between the Nekhoroshev bound and the present study, there are two important differences. First, in the limit of an infinite system, the Nekhoroshev bound is trivially satisfied, while the present effect survives in the limit of $N\to\infty$. Second, Arnol'd diffusion is valid in the entire phase space, while statistical prethermalization occurs only for initial conditions that correspond to low prethermal temperatures.

As mentioned in the introduction, discrete time crystals are novel non-equilibrium phases that do not have any 
static analogues. These phases can be observed when the time-translation symmetry of the periodically driven systems is 
broken spontaneously. Time crystals were, first, predicted theoretically in driven quantum systems and, later, observed in experiments~\cite{zhang17observation,choi2017observation}. Recently, a prethermal time crystal has been observed in a quantum simulator experiment with high frequency 
drive~\cite{kyprianidis2021observation}. In parallel, signatures of discrete time-crystals have been found in classical systems \cite{shapere12classical,yao20classical}. An interesting question for future study is whether classical time crystals can be protected by statistical Floquet prethermalization~\cite{pizzi21classical,pizzi21classical2}.

\section{Acknowledgements}
We acknowledge useful discussions with Itzhack Dana and Roberta Citro. AR would like to thank the department of physics of Bar-Ilan
University for computational facilities. This work was supported by the Israel Science Foundation, Grants No. 151/19 and 154/19. 
AR acknowledges UGC, India for start-up research grant F. 30-509/2020(BSR).

\bibliographystyle{mystyle2}
\bibliography{prethermal}

\end{document}